\documentclass[preprint,prl,aps]{revtex4}
\usepackage{amsmath}
\usepackage{graphicx}
\usepackage{color}
\begin{document}

\title{Surface plasmon based thermo-optic and temperature sensor for microfluidic thermometry}

\author{L.J. Davis III}
\affiliation{Department of Physics, 1274 University of Oregon, Eugene, OR 97403}
\author{M. Deutsch}
\affiliation{Department of Physics, 1274 University of Oregon, Eugene, OR 97403}

\begin{abstract}
We report on a non-interacting technique for thermal characterization of fluids based on surface plasmon resonance interrogation. Using liquid volumes less than 20$\mu$L we have determined the materials' thermo-optic coefficients with an accuracy of better than $1 \times 10^{-5}$ $^{\circ}\mathrm{C^{-1}}$ and demonstrated temperature sensing with an accuracy of $0.03^{\circ}\mathrm{C}$. The apparatus employs a low-power probe laser, requiring only a single wavelength, polarization and interrogation angle for accurate characterization. The device is particularly suited for precise diagnostics of liquids and gases within microfluidic systems, and may also be readily integrated into a variety of lab-on-chip platforms, providing rapid and accurate temperature diagnostics.
\end{abstract}


\maketitle

\section{\label{sec:intro}I. INTRODUCTION}

Accurate and precise knowledge of temperature conditions inside microfluidic systems is crucial for a variety of devices and applications, such as chemical microreactors~\cite{microreactor2} and biofluidic devices~\cite{Rowat}. From the control of micro processor temperature~\cite{Maltezos2006a}, microfluid motion~\cite{Weinert2008}, and biophysical processes~\cite{Kopp1998} to \textit{in situ}~\cite{Valentino2005} sensing for lab-on-chip technologies~\cite{Peterson2002,Stone2004,Fan2007} the needs for non-invasive micro-scale temperature sensing are widespread and growing. However, measurement of temperature on the microliter to nanoliter scale is challenging because standard techniques normally require exchange of heat to achieve thermal equilibrium between probe and sample~\cite{Childs2000}. This leads to the restriction that sample heat capacity be much larger than that of the probe for accurate temperature measurements. While nanoscale metal thermocouples have been recently demonstrated~\cite{Selzer}, their novel architectures may not always be compatible with temperature sensing of fluids either confined to nanoliter volumes or under flow conditions.

Alternatively, using an electromagnetic (EM) field to probe a temperature dependent refractive index allows for highly accurate and non-invasive temperature sensing~\cite{Childs2000}. The thermo-optic (TO) effect, where the refractive index $n$ of a material exhibits dependence on the temperature $T$, is characterized by the thermo-optic coefficient $dn/dT$ and is the origin of thermally induced optical signals. Thus, proper optical monitoring of temperature-related changes in the refractive index of a material with a known TO coefficient enables precise knowledge of a system's temperature. Recently it has been shown that thermo-optical monitoring may be applied to detect the binding of proteins at metal-liquid interfaces~\cite{TOsensor1}.

Accurate measurement of the TO coefficient requires simultaneous knowledge of a material's refractive index and temperature. Various optical methods, such as thermoreflectance measurements~\cite{Fan2000} and the minimal deviation method~\cite{Daimon2007} have been used for measuring the TO coefficient of liquids. However, due to limitations of many experimental setups, measurements of $n$ and $T$ in bulk do not always overlap in space or time. Due to this separation, existing temperature gradients may hinder accurate determination of TO coefficients and additional modeling of thermal transport processes is often necessary in order to correct for this. These additional sources of error may contribute to the spread in reported TO values, which is often greater than 10$\%$, even for well characterized and commonly used materials such as ethanol~\cite{Fan2000}.

We report on the development of an optical reflectance method which employs surface plasmon resonance (SPR) monitoring for accurate determination of the TO coefficients of fluids. This method relies on optical interrogation using a surface-propagating EM wave which is \emph{exponentially} sensitive to changes in the dielectric properties of its environment. Use of SPR monitoring as a sensitive and accurate optical thermometry technique has been studied both theoretically~\cite{Sharma2006,Lin2007} and experimentally~\cite{CHADWICK1993,Chiang2007}. Albeit, the potential of an integrated optical SPR-based sensor for real-time, microfluidic thermal characterization has not yet been realized. We present here a SPR-based system which is capable of TO coefficient determination as well as optical microthermometry. When measuring temperature, the system is readily integrable with lab-on-chip platforms. The apparatus and procedure reported here are specifically designed to mitigate the technical challenges and limitations of TO measurement mentioned above. In particular, by utilizing fluid sample volumes in the micro liter range and embedding the system in a thermal reservoir, thermal gradients are eliminated and equilibration times are significantly reduced, minimizing spatial separation between refractive index and temperature measurements during TO coefficient determination. This allows highly precise measurement of TO coefficients with accuracy better than $1 \times 10^{-5}$ $^{\circ}\mathrm{C^{-1}}$. Such accuracy in TO measurement is becoming increasingly necessary in sensing applications where the TO effect is often a primary source of noise. For example, SPR-based refractive index sensors have recently been shown to have resolution of $1 \times 10^{-7}$ refractive index units (RIU)\cite{homola2008}, making it necessary to control the temperature to within $1\times 10^{-3}$ $^{\circ}\mathrm{C^{-1}}$. However, with the accurate measurement of TO coefficients and temperature enabled by this setup it is possible to correct for the TO-induced noise and eliminate the need for such accurate temperature control.

Once the TO coefficient of the fluid has been determined, the same setup can be used for continuous determination of the sample temperature. This is accomplished by continuous monitoring of the reflectance from the sample at a fixed interrogation angle and a single wavelength. This new \emph{reflectance interrogation} method~\cite{Chen2007} is significantly faster than both angular and wavelength interrogation techniques, which have been usually applied in SPR sensing. Moreover, this now enables SPR-based real-time thermometry in a range of microfluidic systems with only a slight modification to the current setup.

\section{II. SPR-BASED THERMO-OPTIC CHARACTERIZATION}

\subsection{\label{sec:SPsensing}A. Optical Temperature Sensing Using Surface Plasmon Resonance Interrogation}

\begin{figure}[h!]
\centering
\includegraphics[width=80mm]{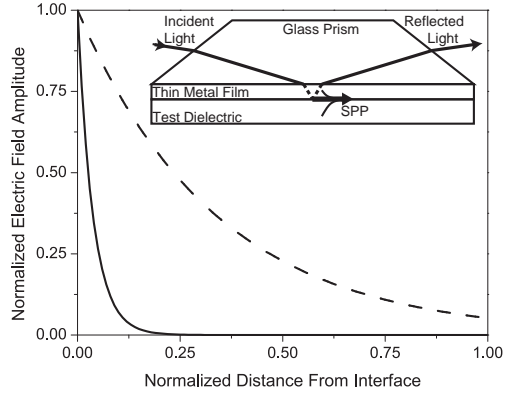}
\caption{\label{fig:1} Normalized amplitude of the SPP electric field, plotted against the distance from the interface normalized to the wavelength, $\lambda = 632.8$nm. The solid line depicts the electric field decay in silver while the dashed line shows the field penetration into water as the adjoining test dielectric. Inset: Schematic (not to scale) of the Kretschmann configuration for excitation of SPPs by $p$-polarized laser light. In this geometry the SPP is excited at the interface of the thin metal film which is not adjacent to the coupling prism, and is thus used to probe the adjoining test dielectric. The dashed lines in the metal film indicate evanescent beams.}
\end{figure}

Surface plasmon polaritons (SPPs) are surface bound EM waves which only propagate when guided along a well-defined metal-dielectric interface. These modes are primarily characterized by a sub-wavelength evanescent confinement of their electric field in directions normal to the interface, resulting in a large EM field enhancement at the interface, as shown in Fig.~\ref{fig:1}. In particular, the electric field component extending into the dielectric medium decays exponentially with a decay coefficient $\gamma_d=\omega\epsilon_d\sqrt{-1/(\epsilon_d+\epsilon_m)}/c$, where $\omega$ is the angular frequency of the incident EM field, $\epsilon_d$ and $\epsilon_m$ are the permittivities of the bounding dielectric and metal, respectively, and $c$ is the speed of light in vacuum. In the visible and infrared EM frequency range the permittivities of metals suitable for the applications discussed here (\emph{\emph{e.g.}} gold, silver) are \emph{negative} and fairly large in absolute value~\cite{PALIK1984}, ensuring the exponential decay of the field into the two bounding media. This decay profile leads to SPPs being highly sensitive to the nanoscale structure and composition of the metal-dielectric interface, thus rendering them excellent probes for surface sensing applications~\cite{HomolaBook}. Typical sensing setups utilizing SPR monitoring rely either on angular or wavelength interrogation of the reflectance in vicinity of the resonance. The best sensors to date are capable of detecting changes of $10^{-7}$ RIU in the test dielectric~\cite{homola2008}.

\begin{figure}[h!]
\centering
\includegraphics[width=80mm]{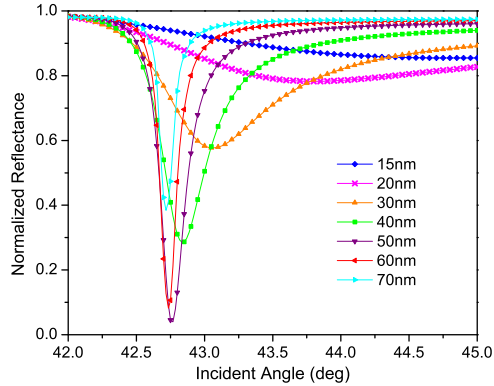}
\caption{\label{fig:2} Reflectance measured as function of incident angle in the Kretschmann configuration, using $p$-polarized light of wavelength 632.8nm and silver films with various thicknesses as stated in the legend. The test dielectric is air at room temperature. The width, magnitude and resonance angle of the reflectance are seen to depend strongly on film thickness, with maximal contrast obtained for film thickness of $\approx50$nm.}
\end{figure}

In all our studies detailed here we utilize the Kretschmann configuration~\cite{RAETHER1988}, shown in the inset to Fig.~\ref{fig:1}. In this configuration a metal film of the order of the optical skin depth in thickness is sandwiched between a dielectric prism and a test dielectric medium with refractive index less than that of the prism. To excite SPPs at the interface between the metal and the test medium, a $p$-polarized, collimated and monochromatic light beam impinges on the metal at an angle greater than the angle for total internal reflection, as shown in Fig.~\ref{fig:1}~\cite{RAETHER1988}. Figure~\ref{fig:2} shows the signature of SPP excitation, evident as a large dip in the reflectance and measured by angular interrogation of the reflected intensity. The angle at which minimal reflectance is observed is known as the \emph{surface plasmon resonance angle}. The SPR angle and corresponding reflectance line shape and width are functions of the permittivities of the metal and test dielectric, as well as the thickness of the metal film, and therefore depend also on the ambient temperature through thermal effects.

\subsection{\label{sec:Calculation}B. Calculated Thermo-Optic Signal}

The origin of the TO effect in most non-conducting dielectrics is due primarily to volumetric thermal expansion. In metals, electron-phonon as well as electron-electron scattering processes also contribute to the variation of the index of refraction with temperature~\cite{Lin2007}. The optical reflectance of a monochromatic light beam with vacuum wavelength $\lambda$ incident at angle $\theta$ at a prism-metal interface as shown in Fig.~\ref{fig:1} can be expressed in terms of the Fresnel reflection coefficients at the prism-metal and metal-dielectric interfaces, $r_{pm}(\theta,T)$ and $r_{md}(\theta,T)$, respectively, as

    \begin{equation}
     R(\theta,T) = \left|\frac{r_{pm}+ r_{md}e^{i2k_{m}h(T)}}{1 + r_{pm}r_{md}e^{i2k_{m}h(T)}}\right|^{2}
    \end{equation}

\noindent Here $h(T)$ is the temperature dependent thickness of the metal film. The wave vector component, normal to the interface, of the light field in the metal layer is given by  $k_{m}=2\pi\sqrt{\epsilon_{m}(T)-\epsilon_{p}\sin^{2}\theta}/\lambda$ and is a function of both $\theta$ and $T$. The arguments in the right-hand-side of Eq. 1 have been omitted for brevity. For simplicity, the permittivity of the coupling prism $\epsilon_{p}$ is assumed to be independent of temperature. While it is straightforward to account for the TO effect in the glass prism, our experimental results indicate that to leading order this is not necessary, since the TO coefficient of the glass used in this setup is more than two orders of magnitude less than that of the test dielectrics used~\cite{Frey}. We further justify this approximation below.

In the linear approximation the temperature dependent refractive index of any non-magnetic medium with permittivity $\epsilon(T)$ is given by $n(T)\equiv\sqrt{\epsilon(T)} = n_0 + (T-T_0)dn/dT$ where $n_0$ is the material-specific refractive index at temperature $T_0$ and $dn/dT$ is the material's thermo-optic coefficient. The linearized temperature-dependent thickness of the metal film is written as $h(T) = h_0[1+\alpha'(T-T_0)]$. Here $h_0$ is the film thickness at temperature $T_0$, and $\alpha'\equiv\alpha(1+\mu)/(1-\mu)$ is a geometric correction to the thermal expansion coefficient of the metal which comes to account for expansion of the metal film mostly in direction normal to the interface, with $\alpha$ and $\mu$ being the metal thermal expansion coefficient and Poisson number, respectively~\cite{Sharma2006}.

\begin{figure}[h!]
\centering	
\includegraphics[width=80mm]{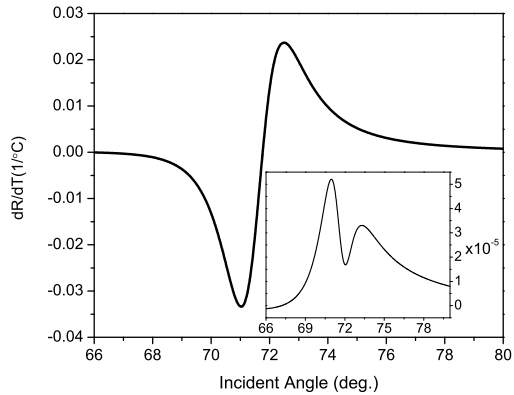}
\caption{\label{fig:3} Temperature derivative of the reflectance plotted against angle of incidence, calculated at room temperature using $\lambda=632.8$nm, $h=50$nm and liquid ethanol as the test dielectric. Inset: Calculated contribution of the metal to $dR(\theta, T)/dT\mid_{T = 22^{\circ}\mathrm{C}}$. Previously published values for the permittivity of silver, $\mu$, $\alpha$ and the TO coefficients of silver and ethanol were used in the calculations~\cite{PALIK1984,Sharma2006,Fan2000}. }
\end{figure}

Using the linearized expressions for $n(T)$ and $h(T)$ we obtain an expression for the temperature dependence of the reflectance written as

    \begin{equation}
    \displaystyle
    \frac{dR}{dT}=\frac{\partial R}{\partial \epsilon_{d}}\frac{\partial \epsilon_{d}}{\partial T}|_{\epsilon_{m},h}+\frac{\partial R}{\partial \epsilon_{m}}\frac{\partial \epsilon_{m}}{\partial T}|_{\epsilon_{d},h}+\frac{\partial R}{\partial h}\frac{\partial h}{\partial T}|_{\epsilon_{d},\epsilon_{m}}
    \end{equation}
\noindent where the left-most term in the right hand side of Eq. (2) denotes the TO contribution from the test dielectric and the remaining terms describe the contributions of the metal film to the measured signal. Figure~\ref{fig:3} shows $dR/dT$ at room temperature ($T_0=22^{\circ}$C) calculated for a 50nm thick silver thin film and liquid ethanol as the test dielectric. The inset to Fig.~\ref{fig:3} shows the calculated contribution from the silver film. We find that the TO contribution from the metal itself is three orders of magnitude smaller than the liquid contribution, therefore allowing us to neglect any TO effects in the metal sensing film in our current experiments. We also calculate the contribution of the glass prism to the TO-induced change in reflectance (using reported values for BK7 glass) and find that it is an order of magnitude less than the contribution from the silver film. Hence the notation in Eq. 2 becomes $dR/dT\approx \left(\partial R/\partial \epsilon_{d}\right)\left(\partial \epsilon_{d}/\partial T\right)|_{\epsilon_{m},h}$.

We see from Eq. (2) that test dielectrics with large TO coefficients will result in greater temperature sensitivity of the device. It is also desirable to utilize test dielectrics with low refractive indices since the SPR reflectance dip broadens with increasing values of the dielectric refractive index, thus leading to overall lower sensitivity. Previous demonstrations of SPR temperature sensors have employed metal-semiconductor junctions exploiting the large TO coefficients of materials such as amorphous silicon. However, most semiconductors exhibit relatively high losses in the optical frequency range as well as large values for their refractive indices, both of which lead to broadening of the SPR and reduced sensitivity~\cite{Chiang2007,CHADWICK1993}. This leads us to conclude that transparent fluids with relatively low refractive indices and large TO coefficients constitute ideal candidates for accurate SPR-based temperature sensing, rendering the latter particularly suitable for applications discussed in the Introduction. Such materials, comprising a large range of liquids and gases also satisfy the approximation discussed above, rendering this setup applicable for accurate thermal characterization in microfluidic lab-on-chip environments. As examples we exploit these properties in ethanol and water to realize a microfluidic, SPR-sensitive TO and temperature sensor.

\section{III. EXPERIMENTAL SYSTEM}

\subsection{\label{sec:Setup}A. Experimental Setup}

\begin{figure}[h!]
\centering
\includegraphics[width=80mm]{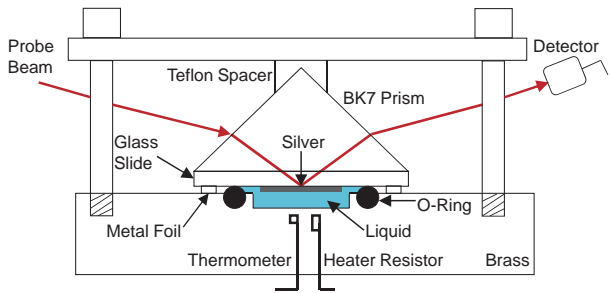}
\caption{\label{fig:4} Diagram of the experimental setup excluding the stages to which the brass-prism system is mounted. The diagram is not to scale.}
\end{figure}

The experimental apparatus for determining TO coefficients consists of a temperature-controlled micro-volume liquid reservoir coupled to an optical monitoring system, shown schematically in Fig.~\ref{fig:4}. A 0.4mm deep reservoir 6.3mm in diameter was drilled into a 1kg brass block which served as the thermal bath. Either water or ethanol were used to overfill the reservoir to a total fluid volume of $\geq12.5\mu$L. To form the optical probing window a 50nm thick silver film 1cm in diameter was thermally deposited onto one side of a glass microscope slide (Corning Glass Works soda lime glass, $n=1.512$) under high vacuum. The opposite side of the slide was index matched to a BK7 glass prism using index matching oil. The prism and slide were then clamped to the brass mass such that the silver film was in direct contact with the liquid. A nitrile o-ring positioned in a shallow groove machined around the fluid reservoir formed a seal between the brass and glass slide. The fractional change in the o-ring cross-section radius necessary to accommodate the liquid thermal expansion in this geometry is of order 0.01, which is well within the material's working tolerances. Metal foil was pressed between the brass and glass slide, around the external perimeter of the o-ring seal to decrease the magnitude of any thermal gradients between the heat bath and the optical interrogation region. Heating was achieved by running electrical current through a 10$\Omega$ resistor embedded within the brass heat bath, about 2cm away from the liquid reservoir. A digital thermometer with $0.01^{\circ}\mathrm{C}$ resolution was also embedded inside the brass at a distance of $250 \mu m$ from the base of the liquid reservoir, and was used to read the temperature of the test dielectric. A collimated, 1mW HeNe laser probe beam ($\lambda =$632.8nm) with 2mm diameter was incident on the silver window through the prism. The brass-prism system was affixed to a rotation stage mounted on a translation stage, allowing for rotation while maintaining the beam spot position centered on the optical interrogation region. A mechanical chopper provided a 911Hz signal modulation of the incoming probe beam, and the reflected signal was detected using a silicon photo diode and a lock-in amplifier.

\subsection{\label{sec:Calibration}B. Thermo-Optic Sensor Calibration}

\begin{figure}[h!]
\centering
\includegraphics[width=80mm]{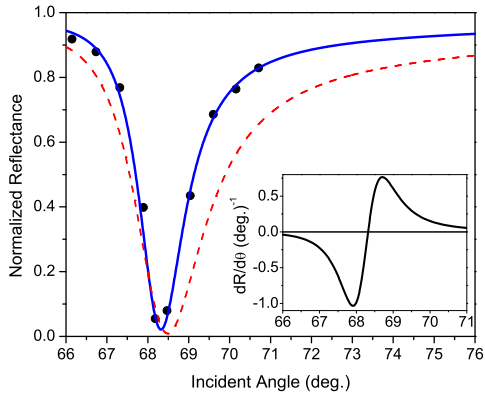}
\caption{\label{fig:5} Reflectance as function of incident angle.  The probe wavelength is 632.8nm, silver film thickness is 50nm and the test dielectric is ultra-pure water (18.2M$\Omega$-cm resistivity) at $T=21.65^{\circ}\mathrm{C}$. Measured values are indicated by dots. The solid curve is the reflectance function obtained by a least-squares fit to the data. The dashed line denotes the calculated reflectance obtained using tabulated values for the refractive index of silver~\cite{PALIK1984}. Inset: Angular derivative of the reflectance calculated using the fitted function, plotted against $\theta$.}
\end{figure}

In order to apply the model discussed in Section II B to our experimental results and measure the liquid TO coefficient it is necessary to have an accurate determination of the refractive index of the silver film. Since the material properties of thin metal films generally depend on parameters such as deposition conditions and ambient environment, we first determine the sample-specific relevant optical constants of our deposited silver films. %however this value can vary between samples and even across a single film~\cite{Barnes2009}.
Using a probe wavelength of 632.8nm we measure $R_c(\theta, T_c)$ -- the reflectance of the silver film as function of incident angle at a known, constant temperature $T_c$. The real and imaginary values of the refractive index, along with the thickness of the film are determined using a least-squares fit of the theoretical reflectance to measured values of $R_c(\theta, T_c)$. We use a DekTak profilometer calibrated to a 3nm height resolution to independently verify the film thickness obtained by the fitting procedure. (While determination of the film's refractive index can also be done using standard ellipsometry, the high sensitivity of the SPR to the polarization of the probe beam eliminates the need for any such additional instrumentation, resulting in comparably accurate values for the optical constants.) The values provided by this procedure are then used to calibrate the sensor. Figure~\ref{fig:5} compares the calculated reflectance for a sample film using measured film thickness and previously reported values for the optical constants of silver~\cite{PALIK1984} with the measured reflectance $R_c(\theta,T_c)$ as well as the corresponding fitted values, illustrating the need for this calibration.

\subsection{\label{sec:Procedure} C. Experimental Procedure and Measurement of the Thermo-Optic Coefficient}

\begin{figure}[t]
\centering	
\includegraphics[width=80mm]{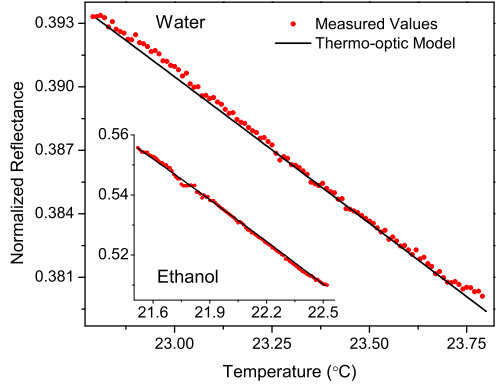}
\caption{\label{fig:6}Reflectance as function of temperature for water (main figure) and ethanol (inset). Measured values are denoted by dots and solid traces indicate the fitted functions.}
\end{figure}

Our first demonstration is of measurements of the TO coefficients of water and ethanol. The rotation stage was initially positioned to produce the desired probe beam-silver film incident angle, $\theta_0$. For maximal sensitivity $\theta_0$ was chosen near the extremal value of $\left|dR(\theta,T)/dT\right|$ as obtained from Fig.~\ref{fig:3}. The temperature of the system was then increased by applying 4 volts across the embedded resistor. When the desired temperature was reached (as determined by reading the embedded thermometer) the heating power was turned off and the system was allowed to relax for 30 seconds. This guarantees that the thermometer readings are consistent with temperature changes of the liquid reservoir, and also serves to eliminate any thermal gradients in the liquid as well as between the liquid and the brass mass. Following this initial relaxation, data acquisition software was used to record the time, temperature, and reflected optical probe signal as the system was cooling, at intervals of $0.01^{\circ}\mathrm{C}$. We note that fixed temperature intervals, equal in magnitude to the resolution of the thermometer were used instead of fixed time intervals when recording data. This is to avoid over-sampling the reflectance and temperature at longer times as the rate of temperature change decreases exponentially during the cooling process. The resulting data set is used to obtain the measured reflectance, expressed as $R_m(\theta_0,T(t))$ with $t$ denoting time. Figure ~\ref{fig:6} shows $R_m(\theta_0,T)$ obtained for water (main figure) and ethanol (bottom inset). Each data point represents an average of ten successive values measured using the lock-in amplifier, with their computed standard deviation used as the measurement error for each averaged point. These errors range in value between $1\times10^{-4}$ and $4\times10^{-4}$, and are not resolved on the scale used in Fig.~\ref{fig:6}.

To model $R_m(\theta_0,T)$ it is necessary to know $\theta_0$ accurately. We see from Fig.~\ref{fig:5} and its inset that the reflectance varies strongly with incident angle in vicinity of the SPR. In fact, $\partial R(\theta,T)/\partial \theta|_{T}$ is large enough near the incident angles of interest that an error in $\theta_0$ of order $\left(1\times 10^{-4}\right)^{\circ}$, well below the angular resolution of standard rotational stages, will introduce a measurable reflectance error of the order of $10^{-4}$. This is a common problem in optical sensors employing angular interrogation, and various subpixelling algorithms have been developed to address it~\cite{Lahav2008}. To enable a more accurate determination of $\theta_0$ in our setup we conduct an iterative least-squares fit of the reflectance, with the TO coefficient and $\theta_0$ as adjustable parameters. Since the reflectance depends strongly on $\theta_0$ we are able to reduce the effective number of fitting parameters from two to one. We find that best fits are obtained when the incidence angle is held fixed at a chosen value and the TO coefficient is varied. Several iterations of this fitting procedure are required, each with a newly adjusted value of $\theta_0$ until a value of the TO coefficient which faithfully reproduces $R_m(\theta_0,T)$ is obtained, for some final value of $\theta_0$ (which is always within the angular measurement error of our system.) The solid lines in Fig.~\ref{fig:6} demonstrate the result of this iterative fitting procedure for both fluids used. We measure the TO coefficients of ethanol and water to be $(4.20\pm0.08)\times10^{-4}$ RIU$/^{\circ}\mathrm{C}$ and $(1.00\pm0.06)\times10^{-4}$ RIU$/^{\circ}\mathrm{C}$, respectively.  These agree very well with the reported values of $(4.10\pm0.47)\times 10^{-4}$ RIU$/^{\circ}\mathrm{C}$ for ethanol~\cite{Fan2000} and $(1.00\pm0.06)\times 10^{-4}$ RIU$/^{\circ}\mathrm{C}$ for water~\cite{Daimon2007,Harvey1998}.

\section{\label{sec:Measurements}IV. EXPERIMENTS AND DISCUSSION}

\subsection{\label{sec:TempSensing}A. Temperature Sensing}

\begin{figure}[b]
\centering	
\includegraphics[width=80mm]{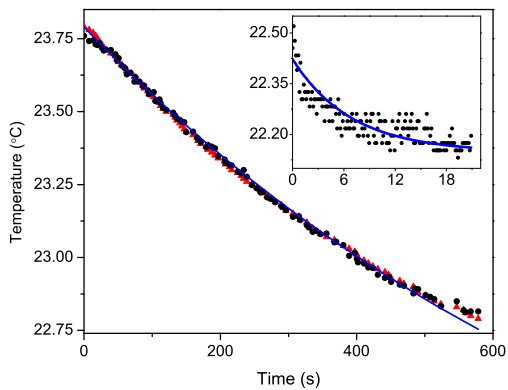}
\caption{\label{fig:7} Temperature plotted against time showing calculated values (black dots) and values measured using the thermometer (red triangles). The blue solid curve denotes a fitted exponential cooling law, with a coefficient of determination $\mathcal{R}^{2}=0.998$. Inset: Non-averaged temperature values obtained from measured real-time reflectance (dots) and fitted exponential cooling law curve.}
\end{figure}

Once $R_m(\theta_0,T(t))$ has been determined for a specific fluid and the TO coefficient is known, it is possible to utilize the same interrogation method for use as an optical temperature sensor. To demonstrate this we apply a linear fit to the data in Fig.~\ref{fig:6}, which we invert to obtain the temperature as function of reflectance and time. Figure~\ref{fig:7} compares $T(t)$, the calculated temperature plotted against time, with temperature values measured by the thermometer during the experiment, demonstrating the accuracy of our method. To determine the measurement error, $\Delta T$, of this temperature sensor we compute $\Delta T = \sigma\left(\partial R_m(\theta_0,T(t))/\partial T\right)^{-1}$ where $\sigma=4\times10^{-4}$ is the measured error in the reflectance. We find $\Delta T$ to range between $0.03-0.06^{\circ}\mathrm{C}$, depending on the sample-specific reflectance line shape and the choice of $\theta_0$.

While we have demonstrated here temperature sensing of a confined and stationary fluid in thermal equilibrium, the system may be easily modified to enable temperature sensing of fluids with known TO coefficients in practical microfluidic systems. In this case the thermal brass mass used for determining the TO coefficient in our setup is not necessary, since the temperature of the fluid is determined by the specific conditions in the flow system. Instead of abutting the fluid reservoir as described previously the prism mounted silver film should be directly contacting the flowing fluid. This can be made possible through a small diagnostics window which may be incorporated into many microfluidic systems~\cite{refractometer}.

As mentioned previously, large errors in measuring the reflectance exist mostly when the system is far from thermal equilibrium and is undergoing rapid cooling or heating. While the signal averaging technique described above is acceptable in near-equilibrium situations when the temperature is varying slowly, it is not practical in cases where rapid changes in temperature are taking place. The calibrated setup used here, however, is accurate enough to allow meaningful temperature measurement in real time using only lock-in detection. The inset to Fig.~\ref{fig:7} shows the non-averaged temperature values plotted against time for rapid cooling of deionized water. The data were obtained by recording the reflectance every 130ms and then converting the data to temperature values using a linear fit of $R_m(\theta_0,T(t))$ data measured according to the procedure in Section III C. We see that even with the existing noise the exponential cooling behavior is still evident, allowing us to extract meaningful temperature values.

\begin{figure}[h!]
\centering	
\includegraphics[width=80mm]{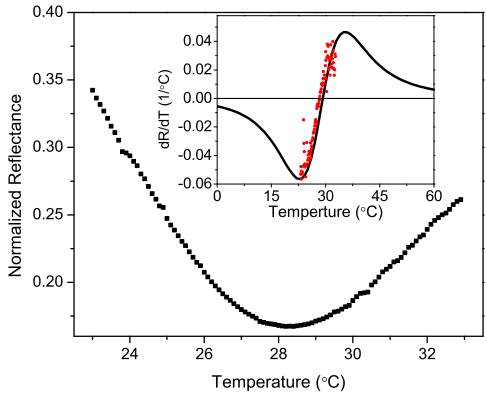}
\caption{\label{fig:8}Reflectance values (raw data) plotted against temperature measured using ethanol as the test dielectric. Inset: Temperature derivative of the reflectance calculated using the theoretical model (solid line) and measured reflectance values from main figure (dots).}
\end{figure}

It can be seen from Fig.~\ref{fig:3} that the magnitude of $\left|dR(\theta,T)/dT\right|$ is significant only over a limited and rather narrow angular range, spanning either side of the resonance angle. The choice of interrogation angle $\theta_0$ to be near the extremal value of $\left|dR(\theta,T)/dT\right|$ guarantees maximal sensitivity for the detection technique used here since it measures changes to the reflectance at the point where its slope is the steepest. However, a large enough change in fluid temperature will shift the SPP resonance angle too far away from the initially optimized value of $\theta_0$, into a lower sensitivity range and potentially even away from the calibration curve of the instrument. The high sensitivity as well as accuracy of this setup are therefore limited to a finite range of temperatures. To verify this we measured the temperature-dependent reflectance over a range of $10^{\circ}\mathrm{C}$, as shown in Fig.~\ref{fig:8}. The inset to Fig.~\ref{fig:8} compares the temperature derivative of the reflectance calculated using our model (solid line) with the data shown in the main figure (dots). We find that $\left|dR(\theta,T)/dT\right|$ drops to half its extremal value over a range of $\sim5^{\circ}\mathrm{C}$ in either direction, and hence set the operational range of this setup to be $10^{\circ}\mathrm{C}$. Since the slope and width of the SPR depend on the permittivity of the dielectric layer, this range will vary with the sensing fluid used.

\subsection{\label{sec:Probe}B. Probe Induced Heating}

Generally the excitation of SPPs in metals is accompanied by damping of the surface-propagating electromagnetic wave, leading to heating of the metal. This heat will be transferred to the liquid reservoir and cause a TO response. It is therefore important to use the lowest probe-laser powers possible while still maintaining sufficiently detectable reflected light powers. Probe-induced heating can be estimated using the steady-state relation $P = \alpha \Delta T_{ss}$ where $P$ denotes the heat transferred into the system from SPP dissipation, $\Delta T_{ss}$ is the temperature difference across the reservoir in steady state and $1/\alpha$ is the characteristic thermal resistance of the system for heat flow across the reservoir (for simplicity we assume one dimensional heat flow and ignore the thickness of the silver film compared to that of the fluid reservoir.) In a standard temperature sensing experiment using water as the test fluid (thermal conductivity = 0.6W/m$\cdot$K) and probe reflectance of about 0.5 we find that the 0.5mW of probe power dissipated in the metal film will result in an increase in steady state temperature of no more than $0.01^{\circ}$. This temperature increase is below the operational resolution of the instrument and will therefore not be measurable under the operating conditions described here. Taking into consideration also the thermal conductivity of the coupling prism, which is typically higher than that of many practical fluids, the rise in temperature in the fluid is guaranteed to remain well below the detection limit. In fact, when using test fluids with lower heat capacity or greater thermal resistance (e.g. ethanol, methanol) it is advantageous to use coupling prisms with large thermal conductivities and low TO coefficients as here. The latter will often have relatively high refractive indices (e.g. BK7 and SF10 glasses,) thus facilitating SPP excitation while minimizing any probe-induced heating effects in the sample.

\subsection{C. Minimal Fluid Volume}

The minimal fluid volume necessary for the SPP-based measurement of TO coefficients and temperature is set by the following two constraints: (i) The liquid-metal interface contact area must be larger than the optical beam spot, and (ii) the fluid thickness (i.e. reservoir depth in our case) must be significantly larger than the decay length of the SPP electric field inside the dielectric medium. In our setup the beam spot at the interface measured less than 6mm in length along its longer axis. Exponential decay lengths of the SPP field depend on the fluid's refractive index, and are typically in the range of 100-300nm. These conditions render nanoliter liquid volumes accessible to this type of temperature measurement. However, such small fluid volumes may be susceptible to probe induced heating and will require proper thermal management as discussed above.

\section{V. SUMMARY}

We have developed a minimally-invasive SPR-based temperature sensor with temperature resolution of $0.03^{\circ}\mathrm{C}$. This sensor and outlined calibration procedure may also be used to obtain the TO coefficients of fluids while addressing typical sources of error in TO coefficient measurement. The apparatus utilizes the exponential sensitivity of surface plasmon polaritons, excited at the interface between a thin silver film and a confined fluid, to extract accurate values of the temperature or the TO coefficient of the fluid. The characterization employs reflectance interrogation, utilizing a low-power probe laser and requiring only a single wavelength, polarization and interrogation angle for accurate determination of temperature. Due to the low refractive indices and high TO coefficients that many fluids possess, this device is particularly suited for precise diagnostics within opto-fluidic systems. Requiring only micro-liter fluid volumes to operate, the setup may be readily integrated into a variety of lab-on-chip platforms, providing rapid and accurate temperature diagnostics.

%\acknowledgments

This work was supported by NSF grant No. DMR-0804433 and ONAMI ONR Grant No. N00014-07-10457. The data in Fig.~\ref{fig:2} were acquired separately by A. Chen.

\end{document}